\documentclass{article}

\usepackage[preprint]{neurips_2024}

\usepackage[utf8]{inputenc} % allow utf-8 input
\usepackage[T1]{fontenc}    % use 8-bit T1 fonts
\usepackage{hyperref}       % hyperlinks
\usepackage{url}            % simple URL typesetting
\usepackage{booktabs}       % professional-quality tables
\usepackage{amsfonts}       % blackboard math symbols
\usepackage{nicefrac}       % compact symbols for 1/2, etc.
\usepackage{microtype}      % microtypography
\usepackage{xcolor}         % colors
\usepackage{graphicx}

\title{Simulating The U.S. Senate: An LLM-Driven Agent Approach to Modeling Legislative Behavior and Bipartisanship}

\author{%
  Zachary R. Baker \\
  University College London\\
  \texttt{zacattackbaker@gmail.com} \\
  \And
  Zarif L. Azher \\
  California Institute of Technology\\
  \texttt{zazher@caltech.edu} \\
}

\begin{document}

\maketitle

\begin{abstract}
This study introduces a novel approach to simulating legislative processes using LLM-driven virtual agents, focusing on the U.S. Senate Intelligence Committee. We developed agents representing individual senators and placed them in simulated committee discussions. The agents demonstrated the ability to engage in realistic debate, provide thoughtful reflections, and find bipartisan solutions under certain conditions. Notably, the simulation also showed promise in modeling shifts towards bipartisanship in response to external perturbations. Our results indicate that this LLM-driven approach could become a valuable tool for understanding and potentially improving legislative processes, supporting a broader pattern of findings highlighting how LLM-based agents can usefully model real-world phenomena. Future works will focus on enhancing agent complexity, expanding the simulation scope, and exploring applications in policy testing and negotiation.
\end{abstract}

\section{Introduction}

Large language models (LLMs) have emerged as powerful computational engines capable of memorization, reasoning, and reflection, using natural language as a medium of input and output. The advent of LLMs – particularly GPT models trained on diverse datasets generated by processing internet content [1] – have generated particular interest in the development of virtual agents. These agents aim to adopt distinct personas according to user design, interacting with other agents and potential actions in sandbox environments to form simulations of scenarios. Such models have reasonably simulated diverse settings, such as hospital management [2], virtual reality games [3], and computer programming [4]. The importance of developing such simulations lies in valuable direct task outputs (ex: automated code generation), as well as the creation of perturbable environments which can be used to better study modeled systems (ex: discovering optimal patient care strategies in hospitals).

The United States Senate is a complex body of governance composed of diverse individuals working to achieve their own agendas. The goal of the U.S. Senate is to propose, debate, and pass legislation that benefits the senators' constituents. The Senate often faces political gridlock and polarization, where neither side is willing to compromise. Efforts to research and better understand political polarization have primarily drawn on social science, including the development of quantitative indicators of political dysfunction as well as takeaways from real-world experience [5] [6].

To date, no study has simulated believable government action using LLMs. We present the first attempt to use LLM-driven virtual agents to simulate the US Senate, focusing on the Select Committee on Intelligence as a proof-of-concept. We seek to demonstrate the potential to use such a system for exploration of legislative behavior and discovery of useful themes, such as how to promote bipartisanship in a gridlocked body.

\begin{figure}
  \centering
    \fbox{%
\includegraphics[width=8cm,keepaspectratio,clip]{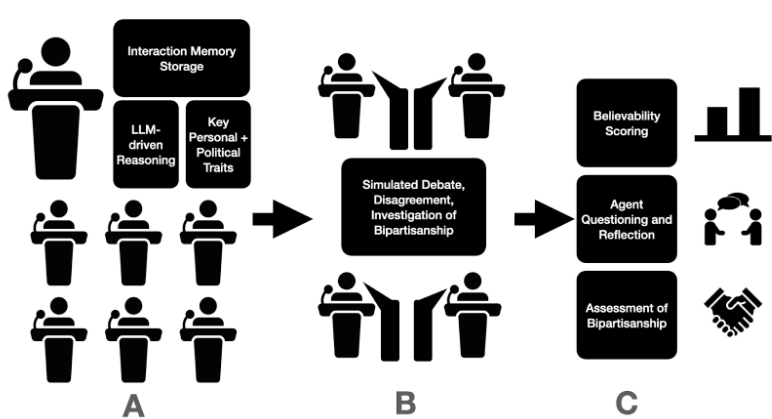}}
  \caption{Study overview. A - LLM-driven agents are synthesized to model U.S Senators. B - Agents are placed into structured debate simulating the Senate Intelligence Committee. C - Agent interactions are evaluated, they are questioned to determine recall and reflection ability, and domain experts assign believability scores to the simulation. }
\end{figure}

\section{Methodology}

We adapted the structure proposed in the seminal “Generative Agents” study [7]. Specifically, we defined agents with specific attributes relevant to their roles as US Senators (name, supported policies, key traits, years of service). Next, we placed them into debate with each other on various complex issues. Each agent was assigned a memory stream which they could refer back throughout debate, to inform discussions with context of previous interactions. At the end of each discussion segment, we could ask the agents to reflect on specific ideas, at which time they accessed the same memory stream to generate relevant summaries.

\subsection{Agents}

Our agents each represented various Senators in the 2024 Senate Committee on Intelligence. To provide accurate information about each Senator to their representative agent, we accessed the popular GPT-3.5 LLM model [8] using the OpenAI API to generate concise policies and key traits about each member of the committee. For instance, for Republican Senator Marco Rubio from Florida, the LLM generated the following policy bio: "My policies are aimed at advancing economic growth and strengthening America's role in the world".

Next, for each Senator, their name, years of service, party, traits, and policies were stored in a JSON file to be accessed during the simulation. Due to resource constraints associated with LLM simulation, we limited the Senators included in the simulation to: Mark Warner (D), Marco Rubio (R), Susan Collins (R), John Cornyn (R), Ron Wyden (D), and Martin Heinrich (D).

Each agent was instantiated with a memory list, which is a JSON that stores the agent's perceived memories at every time step. At each step, their interpretation of the current conversation was stored for accessed later in the simulation. Throughout debate, the agents accessed their memories to make more informed comments in the context of previous actions. At the end of our simulation, we questioned the agents and asked them to reflect on the debate and what was accomplished using the same memory list.

\subsection{Simulation}

We presented our agents with two different current issues to discuss: 1) Russia's invasion of Ukraine. 2) A general discussion of ideas for a bill that the nation needs. For each topic, we presented the agents with a prompt describing the problem settings. Then, they each had an opportunity to state their position. During each cycle, agents could respond to others as they saw fit, resulting in both cooperation and arguments. After three rounds of such conversation cycles, we began questioning and evaluation.

\subsection{Assessment of agent capability and simulation believability}

Our evaluation focused on two key aspects of the simulated Senate Intelligence Committee: (1) the agents' ability to summarize and reflect on their actions and (2) their capacity to find bipartisan solutions under varying conditions. These goals aimed to assess the efficacy of our LLM-driven virtual agents to simulate realistic senatorial behavior and decision-making processes.

To assess agent capacity for self-reflection and summarization, we randomly selected agents from our simulations and prompted them to reflect on their actions during committee meetings. At the end of each conversation cycle for each presented problem, we asked our chosen agents to reflect on the debate experience and what they accomplished, through three questions:  “What did you do during committee?”, “What senator did you agree most with, and which did you disagree most with?” and “What progress was made today, and what held you back the most?” We scrutinized their responses for accuracy, depth of insight, and alignment with their simulated behavior.

For quantitative analysis further we recruited two domain experts – a high school Government teacher and a former Congressional staffer –  to assign believability scores (0 = not believable, 10 = highly believable) to the generated simulations. For this analysis, the entire simulation was run 10 times, with experts assigning a single score to each round. Inter-observer agreement was assessed using Pearson’s correlation testing.

To assess the ability of our simulation to illuminate pathways to bipartisanship, we introduced perturbations to each scenario to potentially catalyze bipartisan cooperation. Subsequent interactions were observed and qualitatively analyzed.

\section{Results}

\subsection{Agents accurately and thoughtfully summarize simulations}

We generally observed that agents demonstrated high accuracy and thoughtful reflections when summarizing their actions. For example:

Agent Rubio: "During the committee meeting, I strongly advocated for substantial military aid to Ukraine. With the new information that Russia is about to overrun Ukraine, I emphasized the urgency of our support. I believe we can't afford to wait, and our response needs to be swift and decisive. I'm satisfied with the outcome, especially since we managed to agree on an emergency aid package that includes both military and humanitarian assistance, along with provisions for oversight."

This reflection accurately captured the agent's role in the debate, primary arguments, and satisfaction with the outcome, demonstrating an understanding of situational urgency and the compromise reached.

Similarly, Agent Wyden provided a nuanced reflection that acknowledged both the urgency of the situation and concerns about accountability:

Agent Wyden: "In the meeting, I raised concerns about the balance between our domestic needs and the urgency of supporting Ukraine. While I understand the gravity of the situation, I stressed the importance of having clear accountability and oversight to prevent misuse of funds. Given the immediate threat, I supported the emergency aid package but insisted on the need for strict oversight and a long-term strategy. I'm cautiously optimistic about the outcome; it's a necessary step, but we must follow through with diligence."

\subsection{Expert-assigned believability}

Domain experts ranked tested simulations as relatively highly believable on a scale from 0-10, with mean scores greater than 5 across scenarios (Table 1). Importantly, inter-observer agreement measured by Pearson’s correlation demonstrated significant (p<0.5) agreement between the experts, strengthening the reliability of the measures.

\begin{table}
  \caption{Mean believability score and calculated inter-observer agreement for agent simulations }
  \label{sample-table}
  \centering
  \begin{tabular}{lll}
    \multicolumn{2}{c}{}                   \\
    \cmidrule(r){1-2}
    Funding for Ukraine     &  \\
    \midrule
    Expert 1 (HS Government Teacher) & 8.1 \\
    Expert 2 (Former Congressional Staffer)     & 6.8 \\
    Pearson’s Correlation, P-value     & 0.63, 0.03 \\
    \midrule
    Discussion on needed bills     &  \\
    \midrule
    Expert 1 (HS Government Teacher) & 6.4 \\
    Expert 2 (Former Congressional Staffer)     & 7.2 \\
    Pearson’s Correlation, P-value     & 0.59, 0.02 \\
    \bottomrule
  \end{tabular}
\end{table}

\subsection{Simulation perturbations identify potential for bipartisanship}

We observed changes in agent behavior and decision-making following introduced perturbations, sometimes leading to bipartisanship. For example:

Initial Debate: Agents demonstrated significant polarization regarding funding for Ukraine. Agents Rubio and Cotton advocated for immediate and substantial military aid, while Agent Wyden expressed concerns about balancing domestic needs with international commitments.

Perturbation: Introduction of intelligence indicating imminent Russian overrun of Ukraine.

Outcome: The perturbation led to increased bipartisanship. Agents who were initially hesitant, such as Wyden, agreed to support an emergency aid package. The committee reached a consensus on providing both military and humanitarian aid, with provisions for oversight.

In Figure 1, we highlight representative dialogue from each phase in this flow of debate. As presented, agents appeared more likely to strike conciliatory tones open to compromise following the introduction of a perturbation.

\begin{figure}
  \centering
    \fbox{%
      \includegraphics[width=10cm,clip]{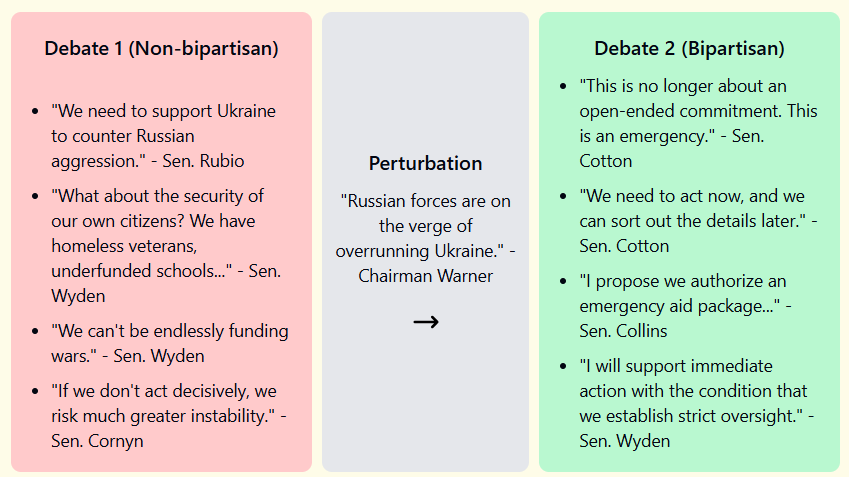}}
  \caption{Introduction of perturbation leads to agent bipartisanship.}
\end{figure}

\section{Discussion}

Here, we demonstrate the potential of LLM-driven virtual agents to simulate complex legislative processes in the U.S. Senate Intelligence Committee. The ability of the agents to provide accurate summarizations and thoughtful reflections suggests that our model can capture nuanced decision-making processes. Expert-assigned believability indicates that outputs from our system are reasonably reflective of real-world Senate proceedings, highlighting the applicability of the framework. The capacity of the simulation to model bipartisanship, especially in response to external perturbations, is significant. This suggests our model could be valuable for studying factors that promote bipartisanship in legislative bodies. Broadly, our findings showcase the utility of using LLM-based virtual agents to improve human understanding of critical real-world scenarios.

Our results align with emerging literature illuminating how LLMs can be valuable tools for enhancing policymaking by helping refine and target legislation [9]. Recently, Moghimifar et al [10] utilized LLMs to model consensus-finding negotiations between political coalitions. However, their methodology did not adapt agent models equipped with memory capability, and their simulation structure utilized Markov Models rather than open-ended discussion between agents. We believe that the application of these differences enable our approach to more flexibly model government actions at the granular individual actor level.

This study was limited to a single Congressional committee with selected members due to computational resource constraints associated with LLM usage. Additionally, we present a preliminary analysis of the proposed methods, where systemic evaluations of adherence to real-world outcomes or comparison of factors driving bipartisanship, were out of scope. Future works can increase agent complexity, expanding the simulation to include the full Congress, and train agents on historical data to simulate past legislative sessions. The proposed framework could further be used to simulate outcomes of legislation, helping policymakers anticipate challenges and proactively identify grounds for bipartisanship.

\section{Conclusion}

Our study presents a novel approach to simulating legislative processes using LLM-driven virtual agents. We demonstrated that these agents can engage in realistic debate, reflect on their actions, and find bipartisan solutions under certain conditions. The ability of our simulation to model shifts towards bipartisanship in response to external factors is particularly noteworthy. As we refine and expand this model, we anticipate it will become a valuable tool for political scientists, policymakers, and educators. By establishing a flexible, controllable environment for studying legislative dynamics, this approach opens up new avenues for research in political science and governance.

\section*{References}

\small

[1] OpenAI (2023). GPT-4 Technical Report.

[2] Li, J., Wang, S., Zhang, M., Li, W., Lai, Y., Kang, X., Weizhi, M., \& Liu, Y. (2024). Agent hospital: A simulacrum of hospital with evolvable medical agents. arXiv preprint arXiv:2405.02957.

[3] Wan, H., Zhang, J., Suria, A.A., Yao, B., Wang, D., Coady, Y., \& Prpa. M., (2024). Building LLM-based AI Agents in Social Virtual Reality. In Extended Abstracts of the 2024 CHI Conference on Human Factors in Computing Systems (CHI EA '24). Association for Computing Machinery, New York, NY, USA, Article 65, 1–7. https://doi.org/10.1145/3613905.3651026

[4] Zhang, K., Li, J., Li, G., Shi, X., \& Jin, Z. (2024). CodeAgent: Enhancing Code Generation with Tool-Integrated Agent Systems for Real-World Repo-level Coding Challenges. arXiv preprint arXiv:2401.07339.

[5] Cameron, M. A., Ribeiro, A., Baier, G., McKay, S., Monnerat, R. A., \& Cameron, C. A. (2022). Partisanship and Political Learning: Lessons from Training Politicians. Journal of Political Science Education, 19(1), 154–173. https://doi.org/10.1080/15512169.2022.2130070

[6] DeFord, D., Dhamankar, N., Duchin, M., Gupta, V., McPike, M., Schoenbach, G., \& Sim, K. W. (2023). Implementing Partisan Symmetry: Problems and Paradoxes. Political Analysis, 31(3), 305–324. doi:10.1017/pan.2021.49

[7] Park, J. S., O'Brien, J., Cai, C. J., Morris, M. R., Liang, P., \& Bernstein, M. S. (2023, October). Generative agents: Interactive simulacra of human behavior. In Proceedings of the 36th Annual ACM Symposium on User Interface Software and Technology (pp. 1-22).

[8] Brown, T. B., Mann, B., Ryder, N., Subbiah, M., Kaplan, J. D., Dhariwal, P., Neelakantan, A., Shyam, P., Sastry, G., Askell, A., Agarwal, S., Herbert-Voss, A., Krueger, G., Henighan, T., Child, R., Ramesh, A., Ziegler, D. M., Wu, J., ... Amodei, D. (2020). Language models are few-shot learners. Advances in Neural Information Processing Systems, 33, 1877-1901.

[9] Peña, A., Morales, A., Fierrez, J., Serna, I., Ortega-Garcia, J., Puente, I., Cordova, J., \& Cordova, G. (2023, August). Leveraging large language models for topic classification in the domain of public affairs. In International Conference on Document Analysis and Recognition (pp. 20-33). Cham: Springer Nature Switzerland.

[10] Moghimifar, F., Li, Y. F., Thomson, R., \& Haffari, G. (2024). Modelling Political Coalition Negotiations Using LLM-based Agents. arXiv preprint arXiv:2402.11712.

\end{document}